\documentclass[10pt,a4paper,oneside]{article}
\usepackage[latin1]{inputenc}
\usepackage[english]{babel}
\usepackage{amsmath}
\usepackage{amsfonts}
\usepackage{amssymb}
\usepackage{graphicx} 
\usepackage{authblk}
\usepackage[multiple]{footmisc}
\providecommand{\keywords}[1]{\textit{Keywords:} #1}

\begin{document}
\title{Quantification of market efficiency based on informational-entropy}
\author{Roland Rothenstein\thanks{Email: roland@rothenstein.org}$^{\ast}$ $\dag$ $\ddag$}

\affil{$\dag$ Norddeutsche Landesbank, D-30159 Hannover\\
$\ddag$ Fachhochschule f\"ur die Wirtschaft Hannover, D-30173 Hannover}

\maketitle

\begin{abstract}
Since the 1960s, the question whether markets are efficient or not is controversially discussed. One reason for the difficulty to overcome the controversy is the lack of a universal, but also precise, quantitative definition of efficiency that is able to graduate between different states of efficiency.
The main purpose of this article is to fill this gap by developing a measure for the efficiency of markets that fulfill all the stated requirements.
It is shown that the new definition of efficiency, based on informational-entropy, is equivalent to the two most used definitions of efficiency from Fama and Jensen.
The new measure therefore enables steps to settle the dispute over the state of efficiency in markets.
Moreover, it is shown that inefficiency in a market can either arise from the possibility to use information to predict an event with higher than chance level, or can emerge from wrong pricing/ quotes that do not reflect the right probabilities of possible events.
Finally, the calculation of efficiency is demonstrated on a simple game (of coin tossing), to show how one could exactly quantify the efficiency in any market-like system, if all probabilities are known.
\end{abstract}



\section{Introduction} \label{sec: Introduction}

Markets are an important part of our daily life. Markets (like supermarkets, the labor market or the stock market) influence nearly every aspect of human interactions. Therefore, their properties are subject to extensive studies. One of the most interesting claims with respect to markets (especially the stock market) is that markets are efficient.
Thus, the so-called 'efficient market hypothesis' (EMH) has become a very powerful and influencing concept. It is, for instance, the conceptional basis for the option pricing theory (see e.g. \cite{Black_1973}) and it supported the invention of exchange traded funds (see e.g. \cite{Gastineau_2002}). 
As a result, the Nobel committee acknowledged work that has been done on the efficient market hypothesis with the Nobel prize 2013 to Fama, Hansen and Shiller \cite{Fama_2013, Hansen_2013, Shiller_2013} 'for their empirical analysis of asset prices'.\footnote{All three laureates have explored different aspects of efficiency from theoretical ground work to analysis of information dynamics in markets \cite{Fama_1965, Fama_1991, Shiller_1981, Shiller_2003, Hansen_1980, Hansen_1991}.}

The surprising decision of the Nobel committee to admit the price to a supporter (Fama) and a skeptic (Shiller) of the EMH, however, shows that, despite of the work that had been done in this field since the 1960 \cite{Beechey_2000}, the question whether markets are efficient or not is still unsolved. On the one hand, is the opinion that markets are overall efficient (see e.g. \cite{Fama_1991, Malkiel_2003a, Malkiel_2003b}), but, on the other hand, exists the view that efficiency is not a good description of market behavior (see e.g. \cite{Shleifer_1990, Shiller_2003}). 
The main purpose of this paper is to overcome the discrepancies between the two positions on the efficiency of markets by developing a measure for the efficiency, as for instance Lo requested it \cite{Lo_1997}. Such a measure of efficiency leads to a new way of looking and analyzing markets. It changes the focus from the question "Are markets efficient?" to the question "How efficient are markets?" It enables the examination of sources of inefficiency and could lead to mechanisms to make markets more efficient. 

The reason why the efficiency of markets is still an unsolved problem lies in various difficulties connected to the problem. One difficulty in the discussion is that the thesis of efficiency as well as the thesis of inefficiency is supported by a wide range of different empirical results. 

The proponents of efficiency studied for instance fund manager performances, examined the random structure of price time series or performed event studies. One of the first was Alfred Cowels. He analyzed that fund manager performances indicate efficiency of markets \cite{Cowles_1933}. Which was further supported by studies of Jensen \cite{Jensen_1968},Treynor
 \cite{Treynor_1965} and Sharpe \cite{Sharpe_1966}. Fama (based on the fundamental work of Samuelson \cite{Samuelson_1965}) examined the random structure of price time series \cite{Fama_1965}\footnote{The random structure of price time series was also independently examined earlier by Mandelbrot \cite{Mandelbrot_1963a}} in context of efficiency of markets.
Furthermore, event studies \cite{Fama_1969, Ball_1968} supported the EMH empirically. 

But, the thesis of inefficiency of markets is also supported by empirical work. 
Starting with studies by Shiller \cite{Shiller_1981} comparing the volatility of markets to that of underlying information and by tests of the ability of markets to represent the information of future dividend structures (see also \cite{LeRoy_1981}), these studies showed evidence that markets are not overall efficient. Further studies showed the inefficiency of markets examining seasonal effects \cite{Ariel_1987}, over-/under-reaction \cite{DeBondt_1985} or earning anomalies, which could be value-effects \cite{Fama_1992}, size anomalies \cite{Banz_1981, Chan_1991} or earning anomalies from technical trading \cite{Allen_1990}. 

This concentration on the verification or falsification of the EMH shows that the current discussion is mainly focused on the binarity of the thesis. Binarity of the thesis means that the EMH is bounded between the extremes of the statements that a market is efficient or inefficient. Even so most researchers know that neither is the case, it is still difficult to find an appropriate description for the state in between.\footnote{In 2004 Lo described an approach to overcome this problem, but he stayed qualitative in his description \cite{Lo_2004}}

In this paper it will be shown that it is possible to overcome this mostly binary discussion over the efficient market hypothesis by deriving a way to measure efficiency gradually. Doing so, it changes the focus of the description from a qualitative to a quantitative definition. The key to overcome the discrepancy in the discussion is to define a universal, mathematical precise definition. 

Another difficulty is based on the proposed definitions for efficiency. 
At the moment, there is no unique, general accepted definition for efficiency. Instead, different definitions were proposed in the past. 
The first article that defines efficiency is by Fama \cite{Fama_1965}. In his work, he summarizes different approaches studying efficiency and he refers to the true value of a stock to define efficiency (see section \ref{subsec: Informational-Efficiency} for a more detailed discussion.)
 In the following, more empirical work on efficiency (as outlined in the previous paragraph) led to a more detailed view on efficiency and an alternative definition by Fama which focused more on information \cite{Fama_1970}. 
A few years later, Jensen \cite{Jensen_1978} emphasized a different aspect with his definition by introducing the impossibility of earning profits as relevant for efficiency (see section \ref{subsec: No-Profit-Efficiency}). In 2004, Timmermann and Granger extended the definition with the aspect of informational costs, search technologies and forecasting models which brings the focus to more practical aspects in the efficiency discussion \cite{Timmermann_2004}. But this extension stays on the same conceptional basis as the definition of Jensen. Therefore, it is not further considered in this paper. Other definitions of efficiency where defined for special purposes, for example in the minority game \cite{Challet_1998}, where efficiency is set equal to a minimum in volatility.


Although, the definitions gave a good intuition about what is meant by efficiency, they were not able to combine mathematical precision with universality. Either the definition is universal, but lacking mathematical precision and manageability, like the first definition of efficiency by Fama (see section \ref{subsec: True-Value-Efficiency}) that is mainly qualitative \cite{Fama_1965}. Therefore, the ability to use it on concrete cases is reduced. Or, the definition has a detailed mathematical description, but it lacks universality, like the Challet-Definition of efficiency on the minority game \cite{Challet_1998}. In this case efficiency can be shown only for special market assumptions. That leads to a situation in which it is ambiguous when a market is called efficient. This paper solves this problem by finding a universal, mathematical precise definition. Furthermore, I show with a theoretical analysis that inefficiency can be caused by two different mechanisms.

The structure of the paper is as follows: In the next section, I describe the status quo of efficiency concepts. In section \ref{sec: Efficiency definition}, I introduce my new efficiency definition which covers the main ideas of the existing definitions (which will be shown in section \ref{sec: Relation to classical approaches}). To derive the new measure, I consider developments on information theory of C.Shannon \cite{Shannon_1948} and C.L.Kelly \cite{Kelly_1956}. Shannon first introduced the concept of information entropy in the context of signal transfer. Kelly used the framework to a game theoretic environment to derive an optimal betting strategy for games. I use this framework to develop my idea of a measurement for efficiency of markets \cite{Rothenstein_2002}.  
 
In section \ref{sec: Sources of inefficiency}, I derive new insights regarding the sources of inefficiency enhancing the definition to pricing mechanisms. The paper concludes by using the new approach to a game of coin tossing, demonstrating how the new approach can work analyzing a real system.

\section{Different views on efficiency} \label{sec: Different views on efficiency}
As described in the introduction, there is no unique, generally accepted definition for the efficiency of a market. Instead, existing definitions are not completely congruent but somehow overlapping and focus on different aspects of efficient markets. The different concepts of efficiency can be associated with different characteristics in the context of efficiency. Without being exhaustive I cluster the different aspects in three approaches.

\subsection{True-Value-Efficiency} \label{subsec: True-Value-Efficiency}
I call the first approach \textbf{true-value-efficiency}. It states that '\textit{actual prices at every point in time represent very good estimates of intrinsic values}'  \cite{Fama_1965}. It means the prices in a stock market reflect at every time the true values of the companies. Unfortunately, the term 'true value' is not a well-defined term in context of efficient markets. One reason might be the model dependency of this approach, (as described in \cite{Shiller_1981} and \cite{LeRoy_1981})  which means that it always leaves the question open if reported inefficiencies are due to a not well defined model or due to real inefficiencies. 
'\textit{Thus, market efficiency perse is not testable.}' \cite{ Fama_1991}

In fact, it is impossible to prove the true-value-efficiency of a system without superior knowledge of the underlying system. In the literature, the superior knowledge is incorporated through valuation models as Fama describes \cite{Fama_1991}. Therefore, efficiency needs a specification in its approach. Such a specification is often strongly connected to assumptions about stock market systems and the valuation of stocks. This means different models could lead to different results about the efficiency of the same market. Since a definition of efficiency should have its focus on a general approach, it is hardly appropriate to use such specific valuation approaches. To define a special model for each new system (for the calculation of a true value) would undermine the intention to define efficiency in a universal way. Therefore, true-value-efficiency is not further considered for a definition of efficiency.

\subsection{Informational-Efficiency} \label{subsec: Informational-Efficiency}
Another characteristic of efficiency is described by Fama as '\textit{a market in which prices always fully reflect available information is called efficient}' \cite{Fama_1970}. I refer to this as \textbf{informational-efficiency}. Informational-efficiency is one of the most referred aspects in the discussion about the efficient market hypothesis (see \cite{Malkiel_1992}). Informational-efficiency claims that any information that influences the price is already priced in the market, because otherwise traders could use this information to generate a profit.\footnote{In later discussions this is weakened by the statement that only information up to a certain information level is incorporated  \cite{Grossman_1980}.} Different kinds of information, which are incorporated in the price, are thereby related to different kinds of efficiency. 

In his famous article from 1970 Fama states three kinds of efficiencies that should be distinguished in the discussion \cite{Fama_1970}\footnote{In a later article, Fama connected these different kinds of efficiency to empirical studies \cite{Fama_1991}. He connected weak efficiency to the test of return predictability, semi-strong efficiency to studies about the influence of new information and strong efficiency to tests of private information.}:
\begin{enumerate}
  \item In a 'weak' efficient market, prices only reflect all information given by past time series.
  \item In a 'semi-strong' efficient market, prices reflect all public available information.
  \item In a 'strong' efficient market, prices reflect all available information (e.g. also insider information).
\end{enumerate}
Throughout this paper, I will refer to strong efficiency (unless otherwise mentioned) to simplify the description of efficiency in the following. By using strong efficiency, a specification of underlying information is not necessary.

Informational-efficiency is a less strong concept of efficiency in comparison to true-value-efficiency, because all true-value-efficient systems are also informational efficient but not vice versa. This can be shown with the following argument: If a true value exists, it clearly includes all information about an asset. So, if the system reflects the true value in the price, it also reflects all information about the asset in the price. Vice versa this is not the case, because it is not mandatory that the true value of an asset must be revealed by all information available.\footnote{Clearly, this argument depends on the interpretation of the term 'true value'. Since the term 'true value' is open for interpretation (as discussed in section \ref{subsec: True-Value-Efficiency}), the conclusion is not the only possible interpretation.}

\subsection{No-Profit-Efficiency} \label{subsec: No-Profit-Efficiency}
An alternative description to the informational-efficiency is given by Jensen: '\textit{A market is efficient with respect to information set $\Theta_t$ if it is impossible to make economic profits by trading on the basis of information set $\Theta_t$.}' \cite{Jensen_1978} 
I will refer to this property as \textbf{no-profit-efficiency}.\footnote{As indicated in the introduction, I interpret $\Theta_t$ in a broad view that includes search technologies and forecasting models as introduced by Timmermann \cite{Timmermann_2004}}
 
No-profit-efficiency means that no market participant is able to outperform the market in a systematic way \cite{Sharpe_1995}.  The definition of no-profit-efficiency follows the same argument as the informational efficiency. Any profit opportunity is already taken before one can use it for a profit in the market.\footnote{Prices in a no-profit-efficient market have to be martingales \cite{Jensen_1978, Samuelson_1965, Mandelbrot_1966}.
A martingale is given by the following formula:
\[E(r)=E(r\vert \phi) \text{   with  } r:\text{price, }  \phi:\text{information}\]
This means that the actual price is the best predictor of the future price independent of information given. Since the concept of martingales is dedicated to time series and not to information in a system, I will not discuss this aspect in greater detail}

The no-profit-efficiency is, however, less strong than the informational efficiency, since all informational efficient markets are also no-profit-efficient but not vice versa. The argument for that is the following: In a market that contains all information, there is no external information on which a profit could be generated. Whereas, for instance, in a market generated by a complex dynamic, which is not foreseeable by traders, also no profit could be generated. But it does not inevitably reflect all information about the system.

\section{Efficiency definition} \label{sec: Efficiency definition}
Based on the efficiency concepts described in the last section, it is now possible to introduce a unique and simple quantitative measure of efficiency, which I will call \textbf{informational efficiency definition}.
Starting with a simplified version of the informational efficiency definition in this section,\footnote{In section \ref{sec: Sources of inefficiency}, I enhance the definition to a more general case.} I show that the main properties of the EMH can be derived from these new definition. In section \ref{sec: Relation to classical approaches}, I show vice versa, that the definition of efficiency can also be derived from the main concepts described in section \ref{sec: Different views on efficiency}. In section \ref{sec: Sources of inefficiency}, it is then possible to introduce an enhanced, more general definition.

Let $X$ be a system with a given probability distribution $p(x)$ for an event $x$ in the system. I define the efficiency of the system $X$ in relation to an information $Y$ as follows:

\begin{equation}
\label{eq:DefinitionEXY}
\text{Eff}(X\vert Y)=\frac{H(X\vert Y)}{H(X)}
\end{equation}

$H(X)$ is the informational entropy of the system $X$ given by\footnote{Following the definition of Shannon \cite{Shannon_1948}}:
\begin{equation}
H(X)=-\sum_x p(x)\log_2 p(x)
\end{equation}

$H(X\vert Y)$ is the conditional entropy of the system $X$ given the information $Y$. For a given information $Y=y$ it is:
\begin{equation} H(X\vert Y=y)=-\sum_y p(y)\bigg( \sum_x p(x\vert y)\log_2 p(x\vert y)\bigg)\end{equation}
$p(x\vert y)$ is the probability that a real event $x$ happens if the information of the event $y$ is given.\footnote{ This probability can be calculated from the probability $p(y \vert x)$ through Bayes formula \[ p(x \vert y) = \frac {p(y \vert x)} {p(y \vert x)+p(\bar{y} \vert x)} \]
where $p(y \vert x)$ is the probability to get the information y if the real event x happens and $p(\bar{y} \vert x)$ is the probability not getting the informations y if the real event x happens.}

$H(X)$ can be interpreted in different ways, which covers different aspects of a system:
\begin{enumerate}
\item It measures the amount of uncertainty in the system.
\item It is the mean informational amount (surprise value) contained in the system. Following Shannon \cite{Shannon_1948}, this is the amount of information that can be transferred.
\item It is a measure of predictability of information content. 
\end{enumerate}

For $H(X\vert Y)$ the following interpretations are common:
\begin{enumerate}
\item $H(X\vert Y)$ describes the uncertainty of the system $X$ given the information $Y$ is known. 
\item The term $H(X\vert Y)$ can show how much predictability is in system $X$ when information $Y$ is known.
\end{enumerate}

Since all interpretations are mathematically identical I will follow all of them in the subsequent sections and refer to the aspect of the entropy that gives the best explanation in each situation. 

Starting with formula (\ref{eq:DefinitionEXY}), I show that the main ideas from the qualitative definition of efficiency can be derived from the new efficiency definition:
Imagine a system is fully efficient and therefore $\text{Eff}(X) = 1$. 
In this case $H(X\vert Y)=H(X)$. The predictability of the system hence is not influenced by an external information $Y$. The efficiency of the system to contain exclusive information and to remain unpredictable is maximal, which corresponds to common qualitative definitions of an efficient market.
 
In case of $\text{Eff}(X)=0$ it follows that $H(X\vert Y)=0$.
The efficiency of the system to contain exclusive information is, in this case, completely inefficient because all information of $X$ could be derived by knowing $Y$, which corresponds to common qualitative definitions of an inefficient market.

Based on the definition, one can further see that any system with $0>\text{Eff}(X) > 1$, hence, $H(X)>H(X\vert Y)$ is not totally efficient. This is plausible since at least some information of the system $X$ can be predicted by using information $Y$. The grade of efficiency in such a case can be quantified between 0\% and 100\% which enables a graduation of efficiency of the system.\footnote{The graduation in efficiency should not be mixed up with the efficiency of a system based on different information sets as Fama introduced them (see section \ref{subsec: Informational-Efficiency}). For the relation between these both see section \ref{subsec: Informational-Efficiency2}.}

As standard efficiency $\text{Eff}(X)$ of a system, I define the special case with the information set $Y^*$:
\begin{equation}
\label{eq:DefinitionEX}
\text{Eff} (X)=\text{Eff} (X\vert Y^*)=\frac{H(X\vert Y^*)}{H(X)}
\end{equation}

Where $Y^*$ represents all available information about the system $X$.
'All' information means the best possible information given to a certain moment in the given environment. This definition of $Y^*$ emphasizes a similarity to the definition of strong efficiency used by Fama (see section \ref{subsec: Informational-Efficiency}).

\section{Relation to classical approaches} \label{sec: Relation to classical approaches}
In this section, I show how the new definition can be derived from the two different concepts for EMH introduced in section \ref{sec: Different views on efficiency}, to demonstrate the universality of the informational efficiency definition.

\subsection{Informational-Efficiency} \label{subsec: Informational-Efficiency2}
The main definition of informational efficiency, first given by Fama \cite{Fama_1970} is:
\begin{itemize}
  \item	\textit{Prices reflect all available information.}
\end{itemize}
To be able to formalize the ideas behind the concept of efficiency to a new definition, I adjust some terms to a more appropriate context. My efficiency definition is not restricted to stock markets; the system to look at could be a system that generates prices (like markets), but it could also be a system that generates quotes (like the coin example in section \ref{sec: Example of tossing coin system}) or it could be any other possible mechanism to make informational content explicit. To take that into account, I substitute the term "prices reflect" by the term "the system contains".  
The term "containing" in the definition means, that information from outside the system cannot change the amount of information of the system $X$. This change in the formulation does not only maintain the essence of Fama's definition of efficiency, but also fulfill the requirement for formalization. With this change, the informational efficiency definition becomes:
\begin{itemize}
  \item	\textit{The system contains all available information.}
\end{itemize}

In the next step, I transfer this formulation into a formula. 
A system $X$ that contains information $Y$ can be described as a system where such an information does not change the informational content of the system given by $H(X)$.
 This can be easily expressed using the mutual information $M(X,Y)=H(X)-H(X\vert Y)$. Because the mutual information is the part of information one could deduce over the system $X$ by knowing information $Y$.
Applying the mutual information to our formulation of efficiency it is described as a system where mutual information 
\begin{equation}
M(X,Y^*)=0
\end{equation}
Where $Y^*$ is the same information used in formula (\ref{eq:DefinitionEX}) in the last section.
$M(X,Y^*)=0$ implies that $H(X|Y^*)=H(X)$. This means the knowledge of $Y^*$ does not change the amount of informational entropy of $H(X|Y^*)$ in comparison to $H(X)$. Thereby all available information $Y^*$ is contained in the system. 

To the contrary, a completely inefficient system is a system where all information of the system $X$ is revealed by knowing information $Y^*$. Such a system is not efficient because the system $H(X)$ contains information that could be completely derived from an information set outside the system. In this case, the conditional entropy $H(X|Y^*)=0$ and the mutual information $M(X,Y^* )=H(X)$. 

Based on these thoughts, $M(X,Y^* )$ can be used to derive a measure of efficiency of a system. To bound the values between zero and one the measure has to be scaled with $H(X)$ (the total information of this system). To get zero as figure for complete inefficiency and one as most efficient figure, the resulting term is subtracted from one.

\begin{equation}
   \text{Eff}(X\vert Y^*)=1-\frac{M(X,Y^* )}{H(X)} =1-\frac{H(X)-H(X\vert Y^* )}{H(X)} =\frac{H(X\vert Y^* )}{H(X)}
\end{equation} 

As a result, I get exactly the same definition as I have defined in section \ref{sec: Efficiency definition}.

So far, I only considered the strong efficiency from the definitions of Fama \cite{Fama_1970}. But semi-strong and weak efficiency can also be taken into account. The connection to semi-strong and weak efficiency can be made by altering the prior knowledge $Y$.  In this case, efficiency is given by $\text{Eff}(X\vert Y_{semi} )$ and $\text{Eff}(X\vert Y_{weak} )$, respectively. 

	\begin{equation}\text{Eff} (X\vert Y_{weak} )=\frac{H(X\vert Y_{weak})}{H(X)}\end{equation} 
	\begin{equation}\text{Eff} (X\vert Y_{semi} )=\frac{H(X\vert Y_{semi} )}{H(X)}\end{equation}
	\begin{equation}\text{Eff} (X\vert Y_{strong} )=\frac{H(X\vert Y_{strong})}{H(X)}=\frac{H(X\vert Y^*)}{H(X)}=\text{Eff}(X)\end{equation}

In contrast to the argument for strong efficiency (that the system contains the information already, as discussed above), the efficiency relative to a partial information set $Y$ can have an additional reason. A possible alternative explanation for a partial efficiency measure $\text{Eff}(X\vert Y)$ is that the information is simply not relevant for the information content of $X$. Taking $Y_{weak}$ as an example, it means that a market could either be 100\% efficient because information about past time series is already contained, or it could be because past time series do not influence the system at all.
In any case, $\text{Eff} (X\vert Y_{weak} )$ shows how efficient the system works in relation to the knowledge of $Y_{weak}$.

In addition to the explanations in the last paragraph, one further difference of the new definition, in comparison to the definition given by Fama, is worth mentioning: It is the dependency from the informational basis. Within the new definition framework, it is for instance possible that a system is 100\% efficient with respect to semi-strong information but only 60\% efficient with respect to strong information.
This can be interpreted as some sort of individual efficiency measure. If a participant perceives only a subset of information ($Y\subset Y^*$), it could be that the system is not 100\% efficient in relation to $Y^*$, but on an individual level (in relation to $Y$) it seems 100\% efficient to the participant. Of course, the other way around is not possible. A system that lacks efficiency on a subset of information could never be more efficient if one gets additional information about the system:
\begin{equation}\text{Eff} (X|Y)<1 \text{ with } Y\subset Y^*\Rightarrow \text{Eff}(X\vert Y^*)<1 \end{equation}

\subsection{No-Profit-Efficiency} \label{subsec: No-Profit-Efficiency2}
Another aspect of efficiency is the no-profit-property (see section \ref{subsec: No-Profit-Efficiency}). To formalize this approach, I follow a work done by Kelly \cite{Kelly_1956}, who defined an optimal winning strategy for a participant in a system (given a probabilistic outcome) without the risk of going bankrupt. Kelly showed, that in this case the maximal capital growth 

\begin{equation}
G_{max}=\lim_{n \to \infty} \log \frac{V_n}{V_0} 
\end{equation}
is given by the mutual information 
\begin{equation}
\label{eq:G_max}
G_{max} (X|Y_i )=H(X)-H(X|Y_i)
\end{equation}

This means, that based on the knowledge $Y_i$ of a participant $i$, the individual maximal capital growth is $G_{max} (X\vert Y_i )$. Following the efficiency definition that in an efficient market no profit is possible (see section \ref{subsec: No-Profit-Efficiency}), the result is that $G_{max}$ should be zero in a completely efficient market. In a completely inefficient market the opposite is the case. This means $G_{max}$ should be maximal. Following equation (\ref{eq:G_max}) the maximal profit is equivalent to $H(X)$.
To derive an efficiency measure between zero and one, I normalize $G_{max}$. To get zero as the result for an inefficient system and one as the result for the most efficient system I subtract the normalized $G_{max}$ from one. It follows:
\begin{equation}
\text{Eff}  (X\vert Y_i )=1-\frac{G_{max} (X\vert Y_i)}{H(X)} =1-\frac{H(X)-H(X\vert Y_i )}{H(X)} =\frac{H(X\vert Y_i)}{H(X)}
\end{equation}

Again, I get the informational definition of efficiency. The difference between no-profit-efficiency and informational efficiency is only given by the informational set. The information set $Y^*$ in section \ref{subsec: Informational-Efficiency2} describes all information about the system. Information $Y_i$ in this section refers only to the information of one participant and, therefore, the efficiency of the system with respect to only one participant. 
If however, the knowledge of one participant $i$ is $Y_i=Y^*$ (that means that one participant knows everything about the system), this participant could be seen as some 'all-knowing' (ideal) participant. In this case, no-profit-efficiency and informational efficiency are equivalent. Therefore, the definition of efficiency could also be formulated as the ability of an ideal trader to gain profits in a market as is described in the paper of Rothenstein and Pawelzik \cite{Rothenstein_2005}. As argued in section \ref{subsec: No-Profit-Efficiency}, the no-profit-efficiency is less strong than the informational efficiency. The reason is, that usually no participant has access to all information and is therefore not able to make a profit, even if it is theoretically possible.

\section{Sources of inefficiency} 
\label{sec: Sources of inefficiency}
In the previous section, I showed that a simplified version of the informational efficiency measure covers all aspects connected to classical definitions of efficiency. For further analysis, it is useful to enhance this definition to cover more complex and realistic situations.
With the enhanced definition, it is possible to derive and discuss two main sources of inefficiency in markets.

\begin{figure} [ht!]
\centering
  \includegraphics[width=0.7\textwidth]{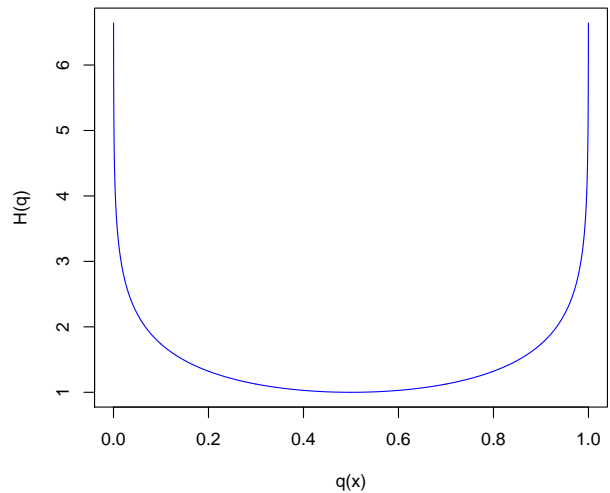}
  \caption{ The graph shows the entropy of a fair coin when the quotes are variable. It shows the dependence from the estimated probability $q(x)$ of an event $x$ determining the quote. The entropy is rising if the quotes are rising. The minimum of the entropy is for a fair probability of the quote ($q(x)=p(x)=0.5$).}
  \label{fig:H_vs_q}
\end{figure}

So far, I have only considered systems in which a given reward is in fair relation to the probabilities of the event. In such a case, the payed reward is $\frac{1}{p}$. This situation is a special case of a more general case where the reward is $\frac{1}{q}$ with $q$ as anticipated probability.\footnote{In this paper, I only consider the case without costs in the system. This means, that $\sum q(x)=1$.} The efficiency definition changes due to the maximal profit one can generate in such a system. 
As Kelly showed \cite{Kelly_1956}, the maximal profit in this case is given by\footnote{In the original paper, Kelly derives his formula not in dependence from the probability $q$ but from the quote $\alpha=\frac{1}{q}$ (fair quote without costs). In this case, the formula is: 
\begin{equation}
G_{max,\alpha} (X\vert Y)=H(\alpha)-H(X\vert Y)
\end{equation}
where
\begin{equation}
H(\alpha)=\sum_x p(x) \log_2 \alpha_x
\end{equation}
with $\alpha$ as the reward payed if the participant predicts the event correct.}

\begin{equation} \label{eq: G_max}
G_{max,q} (X\vert Y)=H(q)-H(X\vert Y)
\end{equation} 
with 
\begin{equation} \label{eq: H_q}
H(q)=-\sum_x p(x)\log_2 q(x) 
\end{equation}

where $p(x)$ is the probability of the event $x$ and $q(x)$ is the anticipated probability to determine the reward. $H(q)$ is the entropy of the system, including consideration for the probabilities of payoff payments. $H(q)$ represents the maximal profit that could be achieved if one has optimal information about such a system. From equation (\ref{eq: H_q}) one could see that $H(q)$ is minimal when the quotes are fair ($q(x)=p(x)$).
It follows, that $G_{max,q} \ge G_{max} \forall q$ and $G_{max,q} = G_{max}$ if $q(x)=p(x)$. 
That means, a player with the optimal strategy can achieve more rewards in a system with unfair quotes than in one with fair quotes (see section \ref{subsec: Fair, unpredictable coin with unfair quotes} and figure \ref{fig:H_vs_q} for an example). 

Taken this into account, the simple efficiency definition: 
\begin{equation}
\text{Eff}(X\vert Y)=\frac{H(X\vert Y)}{H(X)}=1-\frac{G_{max}}{H(X)}
\end{equation}

changes to the efficiency definition including unfair quotes:

\begin{equation}
\text{Eff}_q  (X\vert Y)=1-\frac{G_{max,q}}{H(q)} =1-\frac{H(q)-H(X\vert Y)}{H(q)} =\frac{H(X\vert Y)}{H(q)}
\end{equation} 

Just like in the definition with $p=q$, the efficiency definition is bounded between zero and one. But in the enhanced case, the denominator is H(q).\footnote{From equation (\ref{eq: G_max}) one can see that $G_{max}$ can be maximal $H(q)$ (in the case where $H(X|Y)=0$). Therefore, the maximal entropy H(q) is used as denominator in this definition, since $H(q)\ge H(X) \ge H(X \vert Y) \text{, this ensures that } 0 \ge \text{Eff}_q(X\vert Y) \le 1$.}

With this extension of the definition, it is now possible to find sources of inefficiency.
Therefore, I start with the enhanced definition of efficiency:
\begin{equation}
\label{eq:DefinitionEqXY}
\text{Eff}_q (X\vert Y)=\frac{H(X\vert Y)}{H(q)}
\end{equation}
 Equation (\ref{eq:DefinitionEqXY}) shows that the efficiency is influenced by two different terms. The numerator, $H(X\vert Y)$, describes the possibility to predict the future of the system $X$ by knowing some information $Y$. $H(X\vert Y)$ is maximal (and therefore equal to $H(X)$)  when there is no prediction-power in the information $Y$. In this case, the efficiency is maximal. But, if there is predictability it follows that $H(X\vert Y)<H(X)$ and the efficiency is reduced. The higher the predictability, the lower is $H(X\vert Y)$, and the lower is the efficiency. In the case of a perfect prediction, both $H(X\vert Y)$  and the efficiency would be zero. This represents inefficiency from predictability.

The second term that influences the efficiency is the denominator $H(q)$. As mentioned above, $H(q)$ is minimal when the quotes are fair. For all other quotes, $q(x) \neq p(x)$ is  $H(q)\ge H(X)$. Since $H(X\vert Y)\le H(X)$, it follows that $\text{Eff}= \frac{H(X\vert Y)}{H(q)} \le 1$. 
Therefore, for any case with $q(x)\neq p(x)$ such a system is necessarily inefficient. This is because in case of unfair quotes, the bookmaker is going to pay out more than he earns.
The difference to inefficiency from predictability is that a participant can earn money (in such a system) without any knowledge about the realization of an event (see section \ref{subsec: Fair, unpredictable coin with unfair quotes}). This represents inefficiency from wrong pricing.

Of course, the second source of inefficiency could not in all cases be separated unanimously from the first, since the knowledge of the true distribution often also reveals knowledge of a single event. But, since the prices/quotes can change independently from the realization of one event, this source of inefficiency can be viewed as independent. Overall, the extension to unfair quotes shows that two nearly independent sources influence the efficiency of a system. One source of influence is on the level of real events and their predictability, the second is on the level of the price dynamics/quotes.

\section{Example of a tossing coin} \label{sec: Example of tossing coin system}
In this section, I illustrate the functionality of the informational efficiency definition by using it to determine the efficiency of a game of coin tossing. The example has the advantage that all figures are computable and, therefore, the different consequences are visible. In such a case, it is possible to calculate exactly how efficient a system is.\footnote{The definition can be used for any information processing system where an informational entropy can be calculated (e.g. minority game \cite{Challet_1997} or the seesaw game \cite{Paetzelt_2013a}).}  I show the calculation for different parametrizations: A fair coin with fair odds, a fair coin with unfair odds and an unfair coin with fair odds.

\subsection{Introduction of the game and the parameters} \label{subsec: Introduction of the system and the parameters}
The model works as follows:

A player can bet an amount of one unit on an event $x$ (head 'h' or tail 't' of a coin). Before the player bets, she gets information $y$ about the result. But, this information $y$ is correct only with a certain probability $p(x\vert y)$.\footnote{Alternatively, information could also be based on beliefs of the agent, which are correct with a certain probability $p(x\vert y)$. The mechanism of measuring efficiency is, therefore, not dependent on external information sources.} If she bets correctly, she gets a reward $\alpha=1/q$, otherwise, if she bets incorrectly, she loses her stake.

The following figures are needed to calculate the efficiency of this model: 
\begin{enumerate}
  \item Probabilities for head and tail:
  \begin{equation}
  p(x='t')=1-p(x='h')
  \end{equation} 
These figures describe the probabilities of a coin to show head or tail after tossing.
  \item Probabilities for right knowledge of the result:
\begin{equation} 
p(y='h'\vert x='h')=1-p(y='t'\vert x='h')
\end{equation}
\begin{equation} 
p(y='t'\vert x='t')=1-p(y='h'\vert x='t')
\end{equation}
with x representing the result of the event and y representing the information about the event.
$p(y='h'\vert x='h')$ is the probability that a player gets the information (or has the belief) that the coin shows head if the result of the coin toss is head. $p(y='t'\vert x='t')$ is, accordingly, the same for the information (or belief) and probability for tails.

Since there are no restrictions to the given information of head or tail, the conditional probabilities $p(y='t'\vert x='t')$ and $p(y='h'\vert x='h')$ are completely independent. However, to reduce the number of free parameters, I introduce the restriction to the conditional probability that the information of the player about the results is symmetric.
\begin{equation}
p(y='h'\vert x='h')=p(y='t'\vert x='t')
\end{equation}
This assumes that the player cannot predict the result of tail better than the result of head.


  \item	Probabilities to calculate the quotes:
\begin{equation}q(x='t')\end{equation} is the expected probability for the event tail 't'. From this probability, the quotes are derived via $\alpha(x='t')=\frac{1}{q(x='t')}$.

As mentioned in section \ref{sec: Sources of inefficiency}, in case of no costs the expected probabilities have to sum up to one. Similar to the case of the real event it follows:
\begin{equation} 
q(x='t')=1-q(x='h')
\end{equation}

For the quote follows:
\begin{equation}
 \alpha (x='h')=\frac{\alpha(x='t')}{\alpha(x='t')-1}
\end{equation}
\end{enumerate}
Given the above parameters, three different properties of this system can be distinguished:
\begin{enumerate}
  \item	Fairness of the result
  \item	Predictability
  \item	Fairness of the quote
\end{enumerate}
Each property has a direct relation to the probabilities and vice versa:
\begin{enumerate}
  \item	The coin is fair if $p(x='t')=p(x='h' )=0.5$. In all other cases the coin is rated as unfair.
  \item	The system will be unpredictable if $p(x\vert y)=p(x)$. In the case of a fair coin, this leads to $p(y)=0.5$. In all other cases, there is some sort of predictability in the system.
  \item	Quotes will be called fair if $q(x)=p(x)$ and therefore $\alpha=1/p(x)$. In case of a fair coin, it follows that $q(x)=0.5$ and therefore $\alpha=2$. Otherwise quotes are unfair.
\end{enumerate}

\subsection{Fair, predictable coin with fair quotes} \label{subsec: Fair, predictable coin with fair quotes}
First, I consider the case where the coin and the quotes are fair. In this case, the efficiency of the coin is determined only by its predictability $p(x|y)$.
With the properties from section \ref{subsec: Introduction of the system and the parameters}, the efficiency of the coin can be calculated in dependence from $p(y\vert x)$. Starting with the definition of efficiency: 

\begin{equation}\label{eq:DefinitionEXYdetail}
\text{Eff}(X\vert Y)=\frac{H(X\vert Y)}{H(X)}=\frac{-\sum p(y)\left( \sum p(x\vert y)  \log_2 p(x|y) \right)}{-\sum p(x)  \log_2 p(x)}
\end{equation}

It follows: 

\begin{equation}
\text{Eff}(X\vert Y)=(p(y\vert x)-1)  \log_2 [1-p(y\vert x)] -p(y\vert x)  \log_2 [p(y\vert x)]
\end{equation}

\begin{figure}[ht!] 
\centering
  \includegraphics[width=0.7\textwidth]{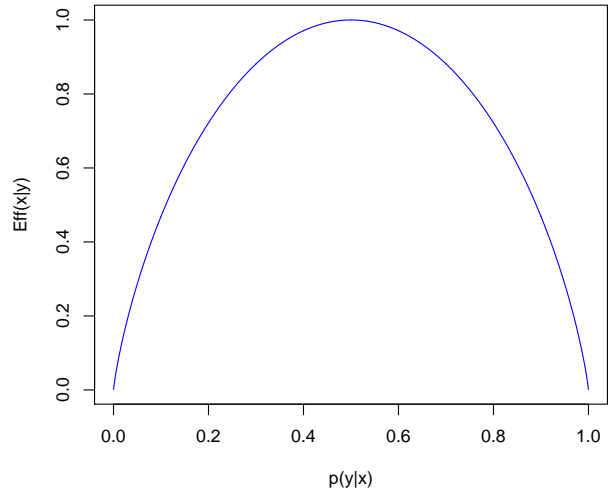}
  \caption{ The graph shows the efficiency of a fair coin in dependence from the conditional probability $p(y \vert x)$. It shows that the efficiency is one when the outcome is not predictable ($p(y \vert x)=0.5$). }
  \label{fig:Eff_vs_prob}
\end{figure}

In figure \ref{fig:Eff_vs_prob}, the curve of efficiency $E(X|Y)$ for a fair predictable coin with fair quotes is shown in dependence from the predictability $p(y|x)$.
The figure shows that the higher the accuracy of the prediction is, the lower is the efficiency. The efficiency ranges from zero (for 100\% and 0\% accuracy of the prediction) to one (if a correct prediction is on chance level). It also shows that the efficiency is still around 50\% even when the outcome can be predicted with nearly 90\%. Furthermore, one can see that a difference of 10\% from unpredictability (that means $p(x\vert y)=0.5\pm 0.05$) influences the efficiency by less than 3\%. However, an inefficiency of 3\% leads in case of coin tossing to 3\% mean revenue per toss, without the risk of going bankrupt.\footnote{This follows from equation \ref{eq: G_max} with $H(X)=1$} 

\subsection{Unfair, unpredictability coin with fair quotes} \label{subsec: Unfair, unpredictable coin with fair quotes}
Second, I consider the case of an unfair, unpredictable coin where the quotes are fair in relation to the probabilities. If we use the respective relations from section \ref{subsec: Introduction of the system and the parameters}, in the efficiency definition shown in formula (\ref{eq:DefinitionEXYdetail}) it follows that the system is always efficient 

\begin{equation}\text{Eff} (X\vert Y)=1\end{equation}

Therefore, in case of unpredictability and fair quotes the efficiency is independent from the fairness of the coin. For instance, if 9 out of 10 times the outcome of the coin is head, but the reward is only 10/9 for this event, it is comprehensibly, that in this case no profit can be generated from the unfairness of the coin. If there is no further knowledge about the occurrence of this event (thus $p(x\vert y)=p(x)$), the efficiency is 100\%. This result illustrates that the efficiency of the coin system does not depend on the fairness of the coin, but on the predictability and the quotes. 

Interesting in this case, is also the informational content $H(X)$ of the system, since it is not static as the efficiency is. $H(X)$ depends on the fairness of the coin in the following way: 
\begin{equation}H (X)=(p(x)-1)  \log_2[1-p(x)]-p(x)  \log_2 p(x)\end{equation} 

\begin{figure} [ht!]
\centering
  \includegraphics[width=0.7\textwidth]{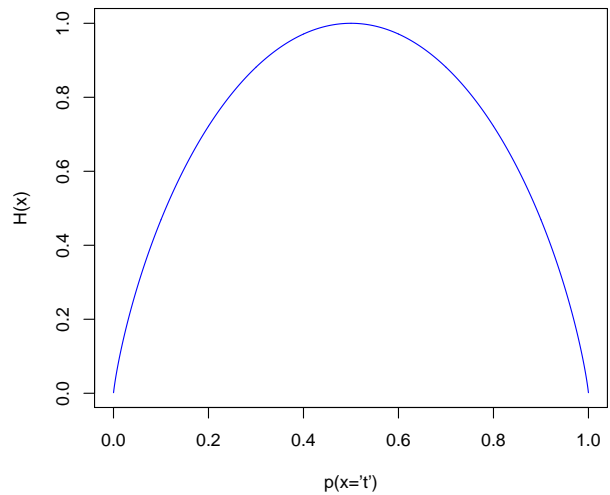}
  \caption{ The figure shows the entropy of the coin system in dependence from the probability of the event showing 'tail'. It shows that the entropy is maximal when the coin is fair and falls to zero if the probability of the coin to show 'tail' is one. The graph for 'head' is identical.}
  \label{fig:H_vs_prob}
\end{figure}

The curve for the dependence of $H(X)$ on the probabilities $p(x)$ is shown in figure \ref{fig:H_vs_prob}. It is easy to understand that a coin that always shows tail $p(x='t' )=1$ has no informational content, and therefore $H(X)=0$. From this point the entropy rises to its maximum, the point where $p(x='t' )=0.5$.

\subsection{Fair, unpredictability coin with unfair quotes} \label{subsec: Fair, unpredictable coin with unfair quotes}

\begin{figure}[ht!]   
\centering
  \includegraphics[width=0.7\textwidth]{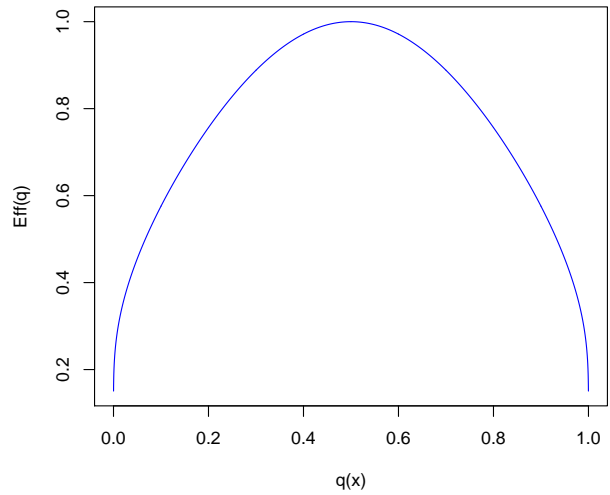}
  \caption{ This graph shows the efficiency of a fair coin that is unpredictable (but has unfair quotes) in dependence from the estimated probability $q(x)$ of an event $x$ determining the quote. It shows that the efficiency is one when the quote is based on the right probability of the event ($q(x)=p(x)=0.5$) and falls to zero if the probability $q$ goes to one and zero, respectively.}
  \label{fig:Eff_vs_q}
\end{figure}

Another interesting case is when the coin is fair and unpredictable but the quotes are not fair. In this case the efficiency (shown in figure \ref{fig:Eff_vs_q}) is:
\begin{equation}E_q (X|Y)=-\frac{1}{0.5 \log_2 q(x)+0.5 \log_2[1-q(x)]}\end{equation}
And for the entropy (shown in figure \ref{fig:H_vs_q}) follows
\begin{equation} \label{eq:H_q_unfair}
H(q)=0.5 \log_2[1-q(x)]-0.5 \log_2 q(x)
\end{equation}

In figure \ref{fig:Eff_vs_q} it is shown that, although there is no predictability for a specific coin toss, the efficiency is not always one. As discussed in section \ref{sec: Sources of inefficiency}, the efficiency of the coin falls if the pricing is not correct even without any predictability of a single event. 



Figure \ref{fig:Eff_vs_q} shows that the efficiency is around 60\% even when the mispricing is about 90\% in comparison to the quote for a fair coin, and that even for a mispricing of 99\% the efficiency is around a relative high value of 30\% efficiency.

\section{Discussion} \label{sec: Discussion}

In this paper, I introduced a definition of efficiency that enhances existing concepts in precision and universality. The new informational efficiency definition enables a substantial step in solving the puzzle, whether markets are efficient or not, by providing a measurement of how efficient markets are. I showed that the new definition based on informational entropy covers the essence of the definition of informational-efficiency (see section \ref{subsec: Informational-Efficiency2}) as well as of no-profit-efficiency (see section \ref{subsec: No-Profit-Efficiency2}). 

By expanding the definition to a wider range of applications (allowing an unfair pricing (see section \ref{sec: Sources of inefficiency}), I showed that inefficiency can have two sources: predictability and unfair pricing. Finally, I demonstrated with a simple system of coin tossing, that, by using the informational efficiency definition, efficiency can be exactly quantified and I showed the efficiency for different cases of a coin. 

One may argue that the new definition only shifts the problem from the uncertainty of the market model to the uncertainty of the information. But this is not the case, as the results of the paper show. The new definition delivers far more advantages: 

First, the new definition is able to derive different grades of efficiency, which was not the case in the previously used definitions. The new definition further provided a new approach to actually measure efficiency quantitatively.
With this approach, the question of whether markets are efficient or not can shift from a fundamental debate to a question of measurement, setting the concept of relative efficiency  \cite{Lo_1997} in place. Therefore, my definition will help to overcome the binary focus in the discussion of efficiency and help to establish a more gradual view. 

Second, I show in section 5 that it is now possible to carry out theoretical analyses based on a clear quantitative definition deriving new insights, like the result that inefficiency can arise from two different sources.

Third, it is now possible to exactly quantify the efficiency for a wide range of toy models.\footnote{The range of toy models is only restricted by the possibility to determine all relevant probabilities (as I showed on a simple model in section 6).}

This introduction is only a first step solving the question of efficiency in markets and further steps have to follow. The time-dependency of the definition can be studied as well as the incorporation of some sort of price dynamic in $q(x)$. Also, I did not consider the case with costs included. But, all these extensions do not change the fundamental concept behind the new efficiency measure.

This fundamental concept of a measure for efficiency shifts the focus of the EMH from a fundamental debate to a question of pure measurement. This transfers the efficient market hypothesis from the actual state of an economic theory into a quantitative science. 

\section*{Acknowledgment}
I would like to thank Franziska Becker, Tobias Basse, Rike Rothenstein and  Wilfried Rickels, for their support in writing this article and Klaus Pawelzik and Mark Kirstein for helpful discussions.

\bibliographystyle{plain}
\bibliography{finance_2018}

\begin{thebibliography}{10}

\bibitem{Allen_1990}
H.~Allen and M.~P. Taylor.
\newblock Charts, noise and fundamentals in the london exchange market.
\newblock {\em The Economic Journal}, 100(400):49--59, 1990.

\bibitem{Ariel_1987}
R.~Ariel.
\newblock A monthly effect in stock returns.
\newblock {\em Journal of Financial Economics}, 18:161--174, 1987.

\bibitem{Ball_1968}
R.~Ball and P.~Brown.
\newblock An empirical evaluation of accounting income numbers.
\newblock {\em Journal of Accounting Research}, 6(2):159--178, 1968.

\bibitem{Banz_1981}
R.~Banz.
\newblock The relationship between return and market value of common stocks.
\newblock {\em Journal of Financial Economics}, 9(1):3--18, 1978.

\bibitem{Beechey_2000}
M.~Beechey, D.~Gruen, and J.~Vickery.
\newblock The efficent market hypothesis: A survey.
\newblock Technical report, Reserve Bank of Australia, 2000.

\bibitem{Black_1973}
F.~Black and M.~Scholes.
\newblock The pricing of options and corporate liabilities.
\newblock {\em Journal of Political Economy}, 81:637--654, 1973.

\bibitem{Challet_1997}
D.~Challet and Y.-C. Zhang.
\newblock Emergence of cooperation and organization in an evolutionary game.
\newblock {\em Physica A}, 246:407, 1997.

\bibitem{Challet_1998}
D.~Challet and Y.-C. Zhang.
\newblock On the minority game: Analytical and numerical studies.
\newblock {\em Physica A}, 256:514, 1998.

\bibitem{Chan_1991}
K.C. Chan and N.~Chen.
\newblock Structural and return characteristics of small and large firms.
\newblock {\em Journal of Finance}, 46(4):1739--1764, 1991.

\bibitem{Cowles_1933}
A.~Cowles.
\newblock Can stock market forecasters forcast?
\newblock {\em Econometrica}, 1(3):309--324, 1933.

\bibitem{DeBondt_1985}
W.~F.~M. {De Bondt} and R.~Thaler.
\newblock Does the stock market overreact.
\newblock {\em Journal of Finance}, 40(3):793--805, 1985.

\bibitem{Fama_1965}
E.~F. Fama.
\newblock The behavior of stock-market prices.
\newblock {\em Journal of Business}, 38(1):34--105, January 1965.

\bibitem{Fama_1970}
E.~F. Fama.
\newblock Efficient capital markets: A review of theory and empirical work.
\newblock {\em Journal of Finance}, 25:383--417, 1970.

\bibitem{Fama_1991}
E.~F. Fama.
\newblock Efficient capital markets: {II}.
\newblock {\em Journal of Finance}, 46(5):1575--1617, 1991.

\bibitem{Fama_2013}
E.~F. Fama.
\newblock Prize lecture: Two pillars of asset pricing.
\newblock www.nobelprice.org, Nobel Foundation, Dec 2013.

\bibitem{Fama_1969}
E.~F. Fama, L.~Fisher, M.~C. Jensen, and R.~Roll.
\newblock The adjustment of stock prices to new information.
\newblock {\em International Economic Review}, 20(1):1--21, Feb 1969.

\bibitem{Fama_1992}
E.~F. Fama and K.~R. French.
\newblock The cross-section of expected stock returns.
\newblock {\em Journal of Finance}, 47(2):427--465, June 1992.

\bibitem{Gastineau_2002}
G.~Gastineau.
\newblock {\em The Exchange-Traded Funds Manual}.
\newblock John Wiley and Sons, New York, 2002.

\bibitem{Grossman_1980}
S.~J. Grossman and J.~E. Stiglitz.
\newblock On the impossibility of informationally efficient markets.
\newblock {\em American Economic Review}, 70(3):393--408, 1980.

\bibitem{Hansen_2013}
L.~P. Hansen.
\newblock Prize lecture: Uncertainty outside and inside economic models.
\newblock www.nobelprice.org, Nobel Foundation, Dec 2013.

\bibitem{Hansen_1991}
L.~P. Hansen and R.~Jagannathan.
\newblock Implications of security market data for models of dynamic economies.
\newblock {\em Journal of Political Economy}, 99(2):225--262, 1991.

\bibitem{Hansen_1980}
L.~P. Hansen and T.~J. Sargent.
\newblock Formulating and estimating dynamic linear rational expectations
  models.
\newblock {\em Journal of Economic Dynamics and control}, 2:7--46, 1980.

\bibitem{Jensen_1968}
M.~C. Jensen.
\newblock The performance of mutual funds in the period 1945-1964.
\newblock {\em Journal of Finance}, 23(2):389--416, 1968.
\newblock May.

\bibitem{Jensen_1978}
M.~C. Jensen.
\newblock Some anomalous evidence regarding market efficiency.
\newblock {\em Journal of Financial Economics}, 6:95--101, 1978.

\bibitem{Kelly_1956}
J.~L. Kelly.
\newblock A new interpretation of information rate.
\newblock {\em Bell System Technical Journal}, 35(4):917--926, July 1956.
\newblock EMH.

\bibitem{LeRoy_1981}
S.~F. LeRoy and R.~D. Porter.
\newblock The present-value relation: test based on implied variance bounds.
\newblock {\em Econometrica}, 49(3):555--574, May 1981.

\bibitem{Lo_1997}
A.~W. Lo, editor.
\newblock {\em Market Efficiency: Stock Market Behaviour in Theory and
  Practice}.
\newblock Edward Elgar Pub, Cheltenham, 1997.

\bibitem{Lo_2004}
A.~W. Lo.
\newblock The adaptive markets hypothesis: market efficiency from an
  evolutionary perspective.
\newblock {\em Journal of Portfolio Management}, 30:15--29, 2004.
\newblock EMH.

\bibitem{Malkiel_1992}
B.~Malkiel.
\newblock {\em Efficient Market Hypothesis}, chapter Vol. I, pages 739--744.
\newblock Macmillan, London, 1992.

\bibitem{Malkiel_2003b}
B.~Malkiel.
\newblock The efficient market hypothesis and its critics.
\newblock {\em Journal of Economic Perspectives}, 17(1):59--82, 2003.

\bibitem{Malkiel_2003a}
B.~G. Malkiel.
\newblock {\em A Random Walk Down Wall Street}.
\newblock W. W. Norton \& Company, New York, 2003.

\bibitem{Mandelbrot_1963a}
B.~Mandelbrot.
\newblock The variation of certain speculative prices.
\newblock {\em Journal of Business}, 36:394--419, 1963.

\bibitem{Mandelbrot_1966}
B.~Mandelbrot.
\newblock Forecasts of future prices, unbiased markets, and "martingale"
  models.
\newblock {\em Journal of Business}, 39(1):242--255, 1966.

\bibitem{Paetzelt_2013a}
F.~Paetzelt and K.~Pawelzik.
\newblock {\em Bubbles, jumps and scaling from properly anticipated prices},
  chapter~5, pages 333--338.
\newblock Springer, Switzerland, 2012.
\newblock e-Print.

\bibitem{Rothenstein_2002}
R.~Rothenstein and K.~Pawelzik.
\newblock Komplexit\"at von aktienzeitreihen in einem evolution\"aren
  aktienmodell.
\newblock In V.~H\"aselbarth, editor, {\em Verhandlungen der DPG (VI)},
  volume~37, page 446. Physik-Verlag GmbH, 2002.

\bibitem{Rothenstein_2005}
R.~Rothenstein and K.~Pawelzik.
\newblock Limited profit in predictable stock markets.
\newblock {\em Physica A}, 348:419--427, 2005.

\bibitem{Samuelson_1965}
P.~A. Samuelson.
\newblock Proof that properly anticipated prides fluctuate randomly.
\newblock {\em Industrial Managment Review}, 6:41--49, 1965.

\bibitem{Shannon_1948}
C.~E. Shannon.
\newblock A mathematical theory of communication.
\newblock {\em B.S.T.J.}, 27(3):379--423,623--656, 1948.

\bibitem{Sharpe_1966}
W.~F. Sharpe.
\newblock Mutual fund performance.
\newblock {\em Journal of Business}, 39(1):119--138, Jan 1966.

\bibitem{Sharpe_1995}
W.~F. Sharpe, G.~J. Alexander, and J.~W. Bailey.
\newblock {\em Investments}.
\newblock Prentice Hall, Englewood Cliffs, N.J., 1995.

\bibitem{Shiller_1981}
R.~J. Shiller.
\newblock Do stock prices move too much to be justified by subsequent changes
  in dividends.
\newblock {\em American Economic Review}, 71(3):421--436, June 1981.

\bibitem{Shiller_2003}
R.~J. Shiller.
\newblock From efficient market theory to behaivoral finance.
\newblock {\em JEP}, 17(1):83--104, Winter 2003.

\bibitem{Shiller_2013}
R.~J. Shiller.
\newblock Prize lecture: Speculative asset prices.
\newblock www.nobelprice.org, Nobel Foundation, Dec 2013.

\bibitem{Shleifer_1990}
A.~Shleifer and L.~H. Summers.
\newblock The noise trader approach to finance.
\newblock {\em Journal of Economic Perspectives}, 4(2):19--33, Spring 1990.

\bibitem{Timmermann_2004}
A.~Timmermann and C.~W.~J. Granger.
\newblock Efficient market hypothesis and forecasting.
\newblock {\em International Journal of Forecasting}, 20:15--27, 2004.

\bibitem{Treynor_1965}
Treynor.
\newblock How to rate management of investment funds.
\newblock {\em Havard Business Review}, pages 63--75, 1965.

\end{thebibliography}

\pagebreak
\onecolumn

\end{document}